\documentclass[conference]{IEEEtran}
\IEEEoverridecommandlockouts
\usepackage{cite}
\usepackage{amsmath,amssymb,amsfonts}
\usepackage{algorithmic}
\usepackage{graphicx}
\usepackage{textcomp}
\usepackage{xcolor}
\usepackage{quantikz}
\usepackage[inline]{enumitem}
\usepackage{orcidlink}
\usepackage{xspace}
\usepackage{microtype}      
\usepackage{physics}
\usepackage{hyperref}

\newcommand{\genpartemail}[2]{\href{mailto:#1@#2}{#1}}

\hypersetup{colorlinks=true,urlcolor=teal, linkcolor=teal,citecolor=teal} 

\def\BibTeX{{\rm B\kern-.05em{\sc i\kern-.025em b}\kern-.08em    T\kern-.1667em\lower.7ex\hbox{E}\kern-.125emX}}
\begin{document}

\title{Trainable Quantum Spectral Models for Partial Differential Equations}

\author{
  \IEEEauthorblockN{
    Gabriel Mejia\IEEEauthorrefmark{1}\orcidlink{0009-0009-8002-1226}, 
    Eileen Kuehn\IEEEauthorrefmark{1}\orcidlink{0000-0002-8034-8837}, 
    Melvin Strobl\IEEEauthorrefmark{1}\orcidlink{0000-0003-0229-9897}, 
    Achim Streit\IEEEauthorrefmark{1}\orcidlink{0000-0002-5065-469X}
  }
  \IEEEauthorblockA{\IEEEauthorrefmark{1}
    Karlsruhe Institute of Technology, Germany,
    \{\genpartemail{gabriel.mejia}{kit.edu}, \genpartemail{eileen.kuehn}{kit.edu}, \genpartemail{melvin.strobl}{kit.edu}, \genpartemail{achim.streit}{kit.edu}\}@kit.edu
  }
}

\maketitle

\begin{abstract}
  This work studies trainable quantum spectral models (QSMs) for solving linear partial differential equations (PDEs).
  Instead of learning solutions directly in physical space, QSMs learn the inverse differential operator in a spectral representation, embedding prior knowledge of the equation's natural basis.

  We systematically study the expressibility and trainability of several QSM architectures, ranging from near-diagonal to fully parameterized unitaries.
  In particular, we introduce a family of richer spectral models that interpolate between purely diagonal operators and fully mixing unitaries through a parameterized mixer controlled by $\epsilon$.
  Our results reveal an intermediate regime, typically around $\epsilon \approx 0.5$, where models achieve the best tradeoff between expressibility and trainability.
  Beyond this threshold, increased circuit complexity degrades convergence without improving accuracy.

  Among the architectures considered, models inspired by the inverse step of the Harrow-Hassidim-Lloyd (HHL) algorithm achieve the fastest training convergence while maintaining high solution fidelity.
  Numerical experiments on the (variable-coefficient) Poisson and Helmholtz equations show that trainable operations in the spectral basis outperform standard variational quantum circuits acting directly in the computational basis.
  These advantages appear through faster convergence, more stable gradients, and more accurate recovery of the reference solution spectrum, particularly through stronger suppression of spurious high-frequency components, even when the operator is not exactly diagonal in the chosen spectral basis.
  Our results identify operator-aware spectral representations as a promising route toward trainable and physically grounded quantum methods for scientific computing.
\end{abstract}


\section{Introduction}

Partial differential equations (PDEs) are fundamental to scientific computing, engineering, and physics, where they describe phenomena ranging from fluid flow and wave propagation to electromagnetism and diffusion.
Classical numerical methods such as finite differences, finite elements, and spectral methods provide accurate and robust solution strategies.
However, their computational cost can become prohibitive for high-dimensional, multiscale, or strongly heterogeneous problems.

Quantum computing has been proposed as a possible route to accelerate the numerical solution of PDEs
\cite{berry_high_2014,berry_quantum_2017,xin_quantum_2020,childs_high_2021,krovi_improved_2022}.
A central line of work relies on methods based on the Quantum Linear Systems Algorithm (QLSA) \cite{harrow_quantum_2009}, which, under restrictive assumptions, may offer substantial asymptotic speedups when combined with efficient state preparation and measurement procedures \cite{aaronson_read_2015}.
Beyond direct linear-system solvers, quantum spectral methods for differential equations
\cite{childs_quantum_2020} further show that structured basis
transforms can yield compact operator representations and more efficient implementations.

In parallel, quantum machine learning (QML) has emerged as a flexible variational framework capable of representing a rich class of functions \cite{schuld_effect_2020,perezsalinas_one_2021}, often associated with rich Fourier or spectral representations of variational quantum models \cite{franz_out_2025,strobl_fourier_2025}.
Variational quantum circuits are especially attractive when exact fault-tolerant algorithms remain impractical.
However, designing trainable quantum models remains a major challenge.
Optimization can be hindered by barren plateaus, noise, and rapidly growing parameter spaces
\cite{mcclean_barren_2018,cerezo_cost_2021,wang_noise_2021,ragone_lie_2024,larocca_barren_2025}.

Several quantum approaches to PDEs have recently been proposed.
Variational formulations \cite{lubasch_variational_2020,liu_variational_2021} and quantum physics-informed neural networks
\cite{kyriienko_solving_2021,panichi_quantum_2025,hunout_variational_2025,cheimarios_solving_2026}
optimize residual or boundary losses directly in physical space.
Other works employ transformed representations or spectral ideas
\cite{leong_variational_2023,pool_nonlinear_2024}. Nevertheless, most existing methods
either focus primarily on algorithmic implementation or rely on generic parameterized circuits, without systematically studying how operator-aligned spectral structure affects trainability and predictive performance.

In many linear PDEs, the governing operator becomes diagonal or low-complexity in a suitable transformed basis.
This suggests that learning in spectral coordinates may be substantially easier than learning directly in physical space.
Importantly, an exact diagonalizing basis is not always required in practice.
For many realistic problems, the underlying physics may be only partially known, coefficients may vary spatially, or no closed-form eigenbasis may exist.
Even in such settings, approximate or physically motivated transformed bases can still expose useful low-complexity structure and improve learning.

Motivated by these observations, we introduce QSMs, a class of variational quantum models that learn inverse differential operators in transformed coordinates.
We compare diagonal, richer spectral, HHL-inspired, fully spectral, and HEA architectures on Poisson, Helmholtz, and variable-coefficient Poisson benchmarks.
Our results show that operator-aligned spectral structure improves convergence, spectral reconstruction, and gradient behavior, even when the chosen basis only approximately diagonalizes the operator.

In this work, we 
\begin{enumerate*}[label=(\roman*)]
  \item introduce QSMs for learning inverse differential operators in spectral coordinates;
  \item compare a hierarchy of architectures to study expressibility, complexity, and trainability;
  \item show that exact diagonalization is not necessary for performance gains;
  \item find that all spectral models, except richer spectral variants with $\epsilon > 0.5$, achieve improved fidelity, convergence, gradient quality, and spectral accuracy.
\end{enumerate*}
In particular, those based on HHL-inspired ansätze provide the best results.

\section{Quantum Spectral Models}

Consider a linear partial differential equation of the form
\begin{equation}
  \mathcal{L}u=f,
  \label{eq:qsm_pde}
\end{equation}
where $\mathcal{L}$ is a linear differential operator, $u$ is the unknown solution, and $f$ is a forcing term or source function.

After spatial discretization, Eq.~(\ref{eq:qsm_pde}) yields the linear system
\begin{equation}
  Au=f,
  \label{eq:qsm_linear_system}
\end{equation}
where $A\in\mathbb{R}^{N\times N}$ is the discrete operator and $u,f\in\mathbb{R}^{N}$.
The exact solution is given by
\begin{equation}
  u=A^{-1}f.
\end{equation}

The central idea of QSMs is to learn this inverse map in a transformed representation where the operator becomes diagonal.
Suppose that $A$ admits a factorization of the form
\begin{equation}
  A = Q \Lambda Q^\dagger,
  \label{eq:qsm_diag}
\end{equation}
where $Q$ is an orthonormal basis transform and
\begin{equation}
  \Lambda=\mathrm{diag}(\lambda_1,\ldots,\lambda_N)
\end{equation}
contains the eigenvalues of $A$. Then the inverse operator can be written as
\begin{equation}
  A^{-1}=Q\Lambda^{-1}Q^\dagger,
\end{equation}
and therefore
\begin{equation}
  u = Q\Lambda^{-1}Q^\dagger f.
  \label{eq:qsm_exact_inverse}
\end{equation}
Equivalently, expanding the solution in the $Q$ basis,
\begin{equation}
  u=\sum_{k=1}^{N}\hat{u}_k \phi_k,
\end{equation}
with spectral coefficients
\begin{equation}
  \hat{u}_k=\frac{\hat{f}_k}{\lambda_k},
  \qquad
  \hat{f}=Q^\dagger f,
\end{equation}
where $\phi_k$ denotes the $k$-th basis vector.

This representation motivates replacing the fixed inverse filter $\Lambda^{-1}$ with a trainable operator acting in transformed coordinates.
We therefore define a QSM as
\begin{equation}
  u_\theta = Q M_\theta Q^\dagger f,
  \label{eq:qsm_general}
\end{equation}
where $M_\theta$ is a parameterized quantum operator and $\theta$ denotes the trainable parameters.

When the transformed basis exactly diagonalizes the operator, the optimal choice satisfies
\begin{equation}
  M_\theta \approx \Lambda^{-1}.
\end{equation}
However, exact diagonalization is not required.
In many realistic applications, the underlying physics may be only partially known, coefficients may vary spatially, or the operator may not admit a closed-form eigenbasis.

In such settings, an approximate or physically motivated basis can still render the operator approximately diagonal, sparse, or compressible, thereby reducing effective complexity and providing a favorable representation for learning.

\subsection{Quantum Implementation}

The right-hand side is encoded into a quantum state,
\begin{equation}
  \ket{f}=\sum_{j=0}^{N-1} f_j \ket{j}.
\end{equation}

Let $U_Q$ denote the unitary implementing the transform associated with $Q$ (for example, a quantum Fourier transform, sine transform, or another structured basis transform).
Applying this unitary gives
\begin{equation}
  U_Q\ket{f}=\sum_{k=0}^{N-1}\hat{f}_k\ket{k}.
\end{equation}

The trainable spectral operator then acts in transformed coordinates,
\begin{equation}
  \ket{\psi_\theta}=M_\theta U_Q\ket{f},
\end{equation}
followed by the inverse transform,
\begin{equation}
  \ket{u_\theta}=U_Q^\dagger\ket{\psi_\theta}.
\end{equation}

Hence, the complete QSM pipeline is
\begin{equation}
  \ket{f}
  \;\longrightarrow\;
  U_Q
  \;\longrightarrow\;
  M_\theta
  \;\longrightarrow\;
  U_Q^\dagger
  \;\longrightarrow\;
  \ket{u_\theta}.
  \label{eq:qsm_pipeline}
\end{equation}

This should be contrasted with conventional variational quantum models, which attempt to learn the full solution map directly in the computational basis,
\begin{equation}
  \ket{f}
  \;\longrightarrow\;
  V_\theta
  \;\longrightarrow\;
  \ket{u_\theta},
\end{equation}
where $V_\theta$ is a generic parameterized circuit.

\section{Model Architectures and Experimental Methodology}

This section describes the specific trainable architectures, benchmark problems, training protocol, and evaluation metrics used in the numerical study.
All models are tested under the same experimental conditions in order to isolate the effect of structural design on expressibility, trainability, and predictive performance.

\subsection{Model Architectures}
While purely diagonal models are well-suited for certain analytically tractable differential equations, they may also admit efficient classical simulation and are not sufficiently general for many realistic physical systems.
We therefore consider a hierarchy of parameterizations for $M_\theta$, ranging from structured low-complexity operators to more expressive quantum circuits.
All QSMs considered here follow the general form
\begin{equation}
  u_\theta = Q M_\theta Q^\dagger f,
  \label{eq:arch_general}
\end{equation}
where $Q$ is the selected basis transform and $M_\theta$ is a trainable operator acting in transformed coordinates.
Different parameterizations of $M_\theta$ define the families of architectures studied below.

\subsubsection{Diagonal Phase Model}

To enforce unitarity while preserving spectral locality, we consider
\begin{equation}
  M_\theta=
  \mathrm{diag}
  \left(
    e^{i\theta_1},\ldots,e^{i\theta_N}
  \right).
\end{equation}

Each transformed mode receives an independent trainable phase.

\subsubsection{Richer Spectral Models}
Purely diagonal spectral models act independently on each transformed mode and therefore cannot represent coupling between spectral components.
To introduce controlled mode mixing while preserving spectral structure, we define richer spectral models composed of alternating trainable diagonal phase layers and fixed mixing blocks:
\begin{equation}
  M_\theta =
  D_{\theta_L} B_\epsilon
  D_{\theta_{L-1}} B_\epsilon
  \cdots
  B_\epsilon D_{\theta_1},
\end{equation}
where
\begin{equation}
  D_{\theta_\ell}
  =
  \mathrm{diag}
  \left(
    e^{i\theta_{\ell,1}},\ldots,e^{i\theta_{\ell,N}}
  \right)
\end{equation}
is a trainable diagonal unitary in the spectral basis.

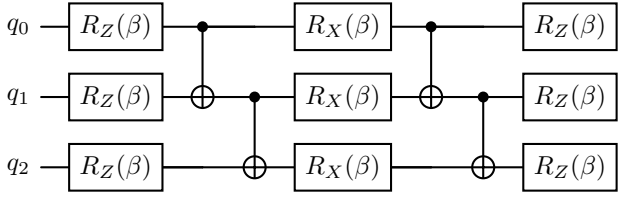
\begin{figure}[t]
  \centering
  \resizebox{0.95\linewidth}{!}{%
    \begin{quantikz}[row sep=0.32cm, column sep=0.38cm]
    \lstick{$q_0$}
    & \gate{R_Z(\beta)}
    & \ctrl{1}
    & \qw
    & \gate{R_X(\beta)}
    & \ctrl{1}
    & \qw
    & \gate{R_Z(\beta)}
    \\
    \lstick{$q_1$}
    & \gate{R_Z(\beta)}
    & \targ{}
    & \ctrl{1}
    & \gate{R_X(\beta)}
    & \targ{}
    & \ctrl{1}
    & \gate{R_Z(\beta)}
    \\
    \lstick{$q_2$}
    & \gate{R_Z(\beta)}
    & \qw
    & \targ{}
    & \gate{R_X(\beta)}
    & \qw
    & \targ{}
    & \gate{R_Z(\beta)}
    \end{quantikz}
  }

  \caption{Compact schematic of the richer spectral architecture
    $B_\epsilon$. Trainable diagonal spectral layers
    \(D(\theta^{(\ell)})\) are interleaved with forward and reverse
    entangling mixers, which introduce controlled coupling between
    spectral modes. The parameter \(\beta(\epsilon)\) controls the
    mixer strength.}
  \label{fig:richer_spectral_B_block}
\end{figure}

The block \(B_\epsilon\) is a fixed unitary mixer whose strength is controlled by a parameter \(\epsilon\in[0,1]\).
In our implementation, the mixing angle is
\begin{equation}
  \beta = \frac{\pi}{2}\epsilon.
\end{equation}
For each mixer block, we apply single-qubit rotations and entangling gates of the form
\begin{equation}
  B_\epsilon =
  \left[
    \prod_q R_Z^{(q)}(\beta/2)
  \right]
  C_{\mathrm{r}}
  \left[
    \prod_q R_X^{(q)}(\beta/2)
  \right]
  C_{\mathrm{f}}
  \left[
    \prod_q R_Z^{(q)}(\beta)
  \right],
\end{equation}
where $q$ is the qubit index, and \(C_{\mathrm{f}}\) and
\(C_{\mathrm{r}}\) denote nearest-neighbor CNOT chains applied in
forward and reverse order, respectively as depicted in Figure
\ref{fig:richer_spectral_B_block}.

Thus, \(\epsilon=0\) gives \(\beta=0\), so \(B_\epsilon\) reduces to the identity and the model remains purely diagonal.
For
\(0<\epsilon<1\), the mixer introduces controlled coupling between
spectral modes.
For \(\epsilon=1\), the mixer reaches its strongest fixed mixing configuration.

This construction allows us to interpolate between a highly structured diagonal spectral model and a more expressive non-diagonal spectral unitary while keeping the number of trainable parameters concentrated in the diagonal layers. From here onward, these are referred to as RS models.

\subsubsection{HHL-Inspired Models}

\begin{figure}[t]
  \centering
  \resizebox{0.95\linewidth}{!}{%
    \begin{quantikz}[row sep=0.32cm, column sep=0.38cm]
    \lstick{$\ket{\hat f_0}$}
    & \ctrl{4}
    & \qw
    & \octrl{4}
    & \qw
    & \gate[wires=4]{Q}
    & \rstick{} \qw
    \\
    \lstick{$\ket{\hat f_1}$}
    & \ctrl{3}
    & \qw
    & \octrl{3}
    & \qw
    & \qw
    & \rstick{} \qw
    \\
    \lstick{$\vdots$}
    & \vdots
    &
    & \vdots
    &
    &
    & 
    \\
    \lstick{$\ket{\hat f_{n-1}}$}
    & \ctrl{1}
    & \qw
    & \octrl{1}
    & \qw
    & \qw
    & \rstick{} \qw
    \\
    \lstick{$\ket{0}_a$}
    & \gate{R_Y(\theta_k)}
    & \push{\ \cdots\ }
    & \gate{R_Y(\theta_k)}
    & \meter{}
    & \qw
    & \qw
    \end{quantikz}
  }
  \vspace{-1mm}
  \[
  \text{post-select ancilla } \ket{1}_a
  \]

  \caption{HHL-inspired model. In the spectral basis, each mode
    \(\ket{k}\) controls a sequence of ancilla rotations
    \(R_Y(\theta_k)\). Post-selection of the ancilla induces the
    learned spectral filtering, followed by the inverse transform \(Q\)
    to recover the solution state in physical coordinates.}
  \label{fig:hhl_free_deep}
\end{figure}
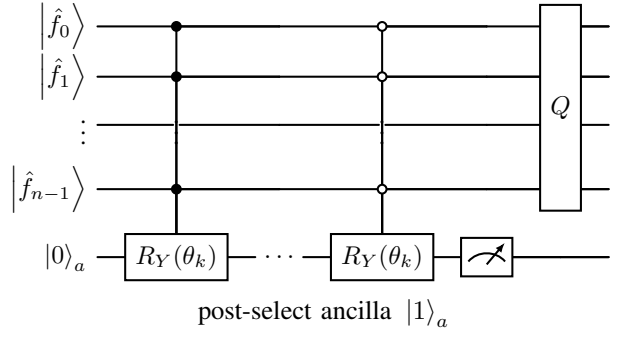

We consider an HHL-inspired architecture based on the inverse-filtering mechanism of the Harrow--Hassidim--Lloyd algorithm (see Figure \ref{fig:hhl_free_deep}), where each transformed mode is assigned an independent trainable controlled-rotation angle:
\begin{equation}
  \alpha_k=\theta_k,
  \qquad k=0,\ldots,N-1.
\end{equation}

The model acts after transformation into spectral coordinates.
For each computational basis state \(\ket{k}\), an ancilla qubit undergoes a mode-dependent controlled rotation,
\begin{equation}
  \ket{k}\ket{0}
  \longrightarrow
  \ket{k}
  \left(
    \cos(\alpha_k)\ket{0}
    +
    \sin(\alpha_k)\ket{1}
  \right).
\end{equation}

Measuring the ancilla and post-selecting the outcome \(\ket{1}\) reweights the amplitudes of the transformed state according to
\(\sin(\alpha_k)\), producing an effective trainable spectral filter:
\begin{equation}
  \hat{u}_{\theta,k}\propto \sin(\theta_k)\hat{f}_k.
\end{equation}

\subsubsection{Fully Spectral Unitaries}

We further consider generic variational circuits acting after the transform,
\begin{equation}
  u_\theta = Q U_\theta Q^\dagger f,
\end{equation}
where $U_\theta$ is a multi-layer hardware-efficient ansatz (HEA) composed of single-qubit rotations and entangling gates.

\subsubsection{HEA Baseline}

As an unstructured reference model, we train a standard HEA directly in the computational basis,
\begin{equation}
  u_\theta = V_\theta f.
\end{equation}

This model learns the full mapping without exploiting operator-aligned transforms.

\subsection{Benchmark Problems}

We evaluate all models on three representative linear PDE families.

\subsubsection{Poisson Equation}

\begin{equation}
  -\Delta u=f.
\end{equation}

This provides an exactly diagonalizable benchmark in sine coordinates.

\subsubsection{Helmholtz Equation}

\begin{equation}
  (-\Delta-\omega^2)u=f.
\end{equation}

This introduces shifted spectra and more challenging inverse behavior.

\subsubsection{Variable-Coefficient Poisson Equation}

\begin{equation}
  -\nabla\cdot(a(x)\nabla u)=f,
\end{equation}
with a spatially varying coefficient, which for all experiments is $a(x) = x$.
This operator is generally not exactly diagonal in the reference basis; therefore, it is tested for robustness under basis mismatch.

\subsection{Dataset Construction}

Training and testing pairs $(f,u)$ are generated synthetically.
Right-hand sides are sampled as low-frequency sine expansions,
\begin{equation}
  f(x)=\sum_{k=1}^{K} c_k \sin(k\pi x),
\end{equation}
with random coefficients $c_k$.

Reference solutions are obtained using classical finite-difference discretizations and direct linear solvers.
In all experiments, the target solutions are constructed using a truncated spectral representation, where only the first $K=7$ modes (out of $N=16$ total modes) carry nonzero energy.
Higher-frequency components are therefore absent in the ground-truth solutions.
This setting reflects the typical smoothness of PDE solutions and provides a controlled scenario to evaluate how different architectures allocate representational capacity across relevant and irrelevant spectral modes.
\subsection{Training Protocol}

All trainable parameters are optimized using the Adam optimizer with a learning rate
$\eta = 10^{-2}$,
which was kept fixed throughout all experiments.
Training is performed for a common epoch budget across all architectures under matched optimization settings whenever possible.

To assess robustness with respect to parameter initialization and stochastic effects, each experiment is repeated over 5 random seeds.
Reported curves correspond to averages across seeds, while shaded regions or error bars indicate one standard deviation.

Unless otherwise stated, all models are trained on the same dataset realizations, with identical train--test splits and comparable parameter budgets whenever feasible.

\subsection{Evaluation Metrics}

We evaluate all models using predictive, spectral, optimization, and expressibility metrics.
Since quantum states are defined up to a global phase, predictions are first aligned with the target via $\phi^\star=\arg(\langle u,u_\theta\rangle)$.

The mean squared error is defined as
\begin{equation}
  \mathrm{MSE}
  =
  \frac{1}{N}
  \left\|
    u-e^{-i\phi^\star}u_\theta
  \right\|_2^2,
\end{equation}
while fidelity is given by
\begin{equation}
  F=|\langle u_\theta,u\rangle|^2.
\end{equation}

In spectral coordinates, with $\hat{u}=Q^\dagger u$ and $\hat{u}_\theta=Q^\dagger u_\theta$, we measure mode-wise reconstruction error
\begin{equation}
  E_k=|\hat{u}_{\theta,k}-\hat{u}_k|^2.
\end{equation}

To characterize trainability, we compute gradient variance across parameters
\begin{equation}
  \mathrm{Var}_g=
  \frac{1}{P}\sum_{i=1}^{P}\mathrm{Var}\!\left(\partial_{\theta_i}\mathcal{L}\right),
\end{equation}
and mode-wise gradient power
\begin{equation}
  G_k=
  \frac{1}{P}
  \sum_{i=1}^{P}
  \left|
    \frac{\partial \hat{u}_{\theta,k}}{\partial \theta_i}
  \right|^2.
\end{equation}

Expressibility is quantified via the Kullback--Leibler divergence between the fidelity distribution induced by random circuit parameters and the Haar reference,
\begin{equation}
  \mathrm{Expr}
  =
  D_{\mathrm{KL}}\!\left(p(F)\,\|\,p_{\mathrm{Haar}}(F)\right),
\end{equation}
where lower values indicate higher expressibility.

\subsection{Objective of the Comparison}

The purpose of these experiments is not only to compare predictive accuracy, but also to understand how operator structure, circuit complexity, and expressibility jointly influence trainability in quantum models for scientific computing.

\section{Results}

We evaluate QSMs on representative linear PDEs and compare them with parameterized HEA.
Beyond predictive accuracy, we analyze trainability, spectral behavior, and the relation between expressibility and optimization.
All results are averaged over multiple seeds, with shaded regions indicating one standard deviation.

\subsection{Overview of Benchmarks}

We consider three problems of increasing complexity: the Poisson equation (exactly diagonalizable in the sine basis), the Helmholtz equation (shifted spectrum), and a variable-coefficient Poisson equation (not diagonal in the reference basis), allowing us to study both ideal and approximate-basis regimes.

\subsection{Training Behavior}

We first analyze training loss and reconstruction error for Poisson and Helmholtz, and report full metrics for the variable-coefficient case.
Since the first two are exactly diagonal in the sine basis, they favor spectral models; the variable-coefficient problem provides a more balanced benchmark.

\begin{figure}[t]
  \centering
  \includegraphics[width=\linewidth]{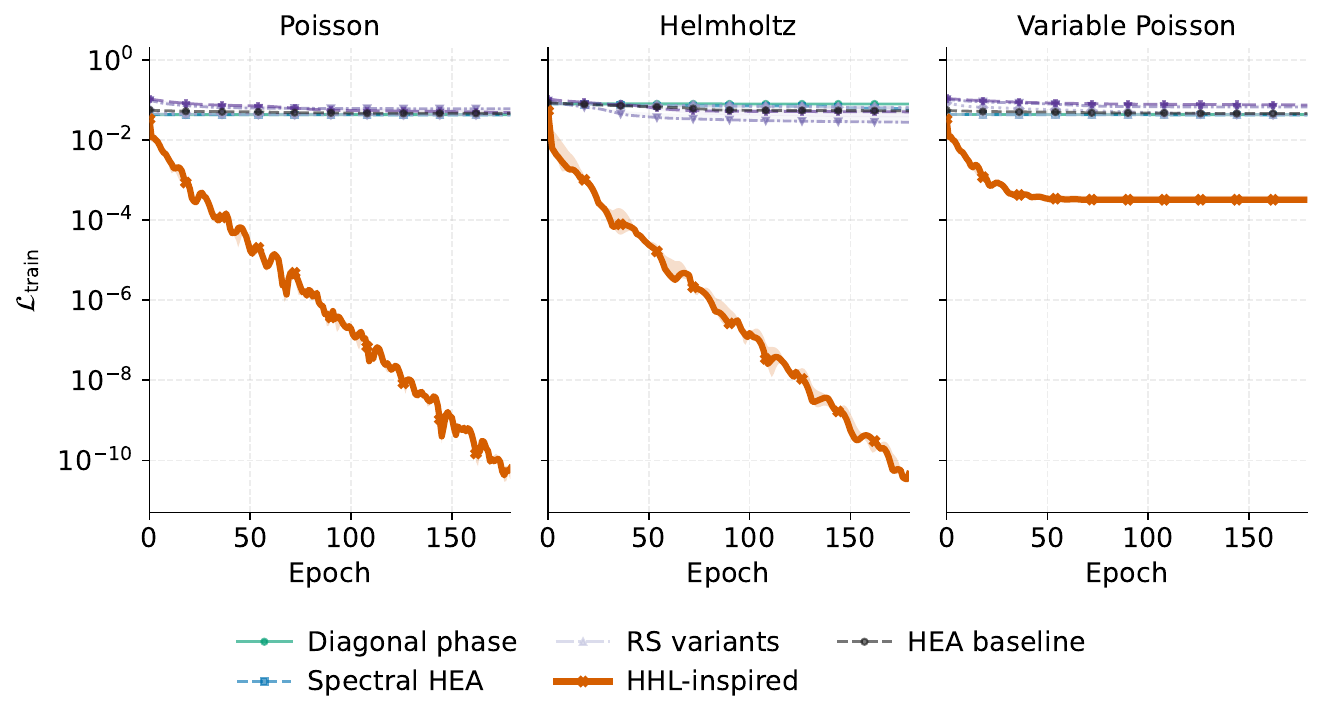}

  \caption{Training loss evolution
    $\mathcal{L}_{\mathrm{train}}=\mathrm{median}(\frac{1}{N}\|u_\theta-u\|_2^2)$
    across PDE benchmarks. All spectral models, except the richer
    spectral model with $\epsilon>0.5$, exhibit faster convergence and
    lower final loss.}
  \label{fig:training_loss}
\end{figure}

Figure~\ref{fig:training_loss} shows that spectral models consistently converge faster than HEA baselines.
Diagonal and HHL-inspired architectures achieve rapid early loss reduction, while more expressive circuits reach comparable performance only with slower or less stable optimization, indicating that expressibility alone does not ensure trainability.

\subsection{Solution Fidelity}

\begin{figure}[t]
  \centering
  \includegraphics[width=\linewidth]{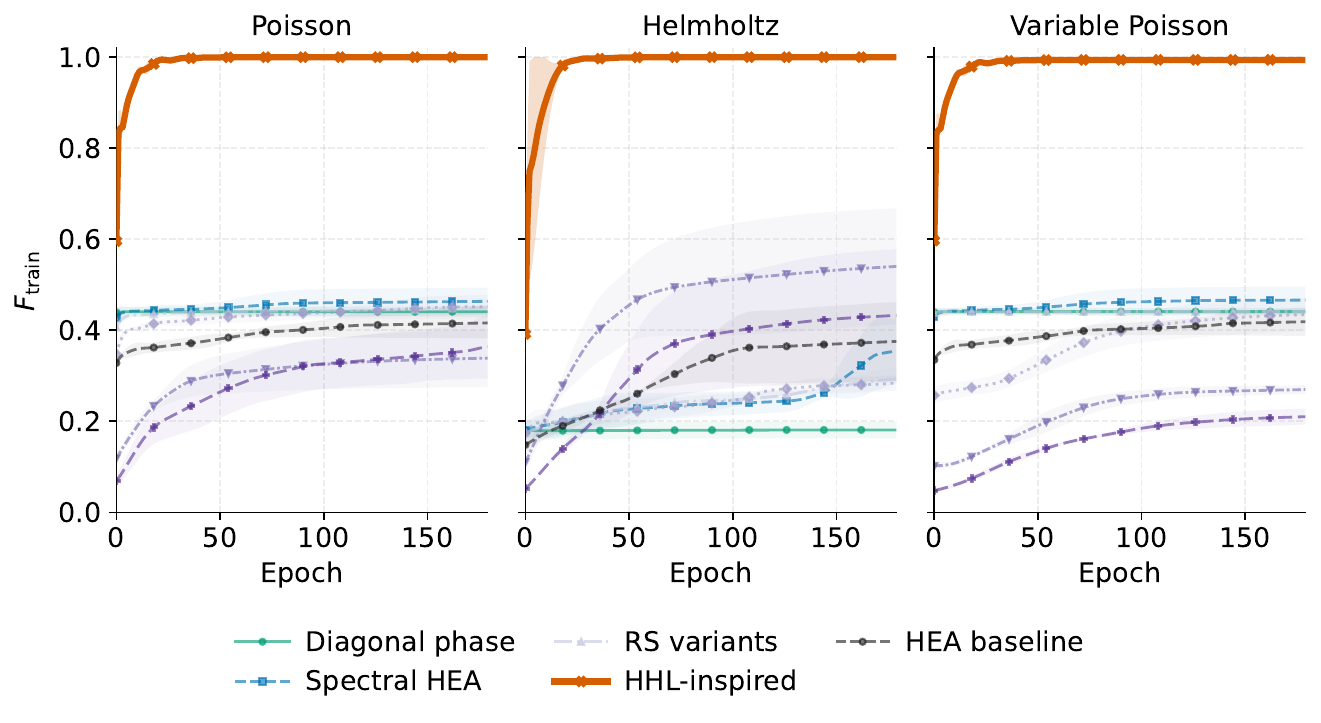}
  \caption{Training fidelity $F=|\langle u_\theta,u\rangle|^2$ for
    all the benchmark PDEs. All spectral models, except the richer
    spectral model with $\epsilon>0.5$, exhibit better fidelity.}
  \label{fig:fidelity}
\end{figure}

As shown in Figure~\ref{fig:fidelity}, QSMs reach higher fidelity more rapidly than generic circuits, with HHL-inspired models combining stable optimization and accurate reconstruction.

\subsection{Spectral Reconstruction}

\begin{figure}[t]
  \centering
  \includegraphics[width=\linewidth]{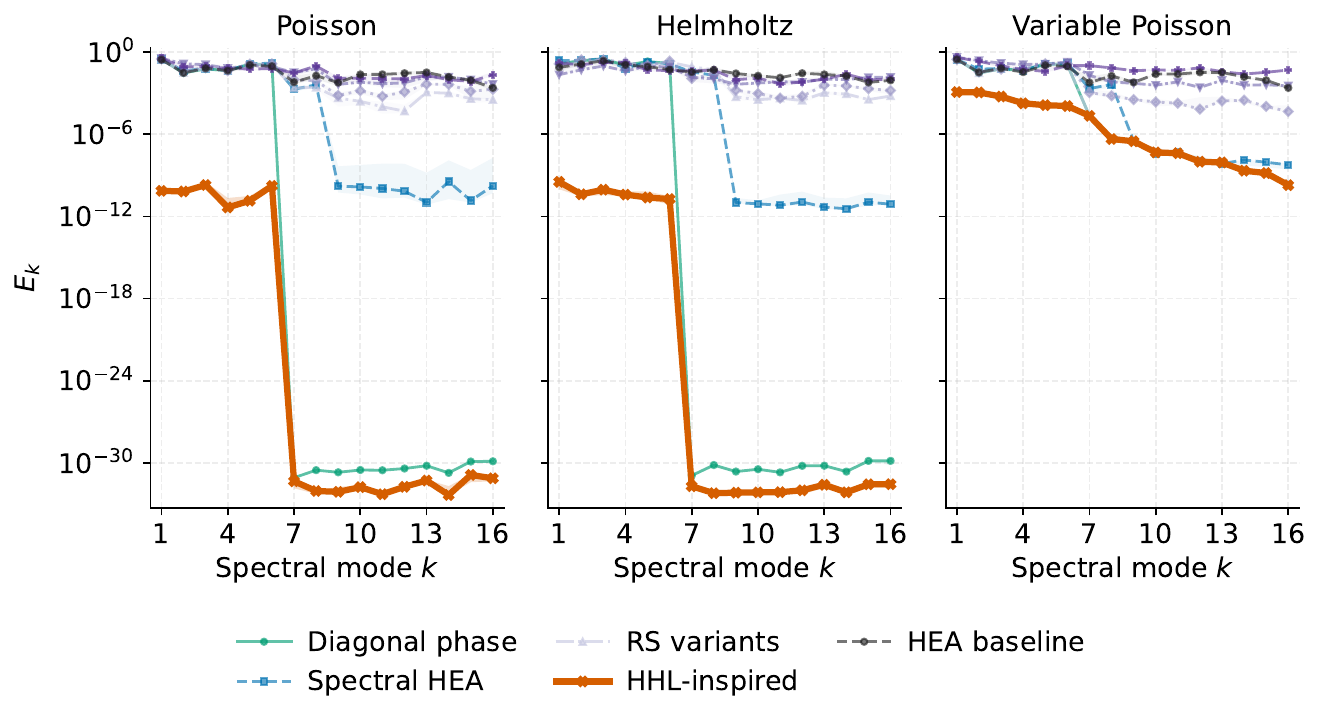}

  \caption{Mode-wise spectral error
    $E_k=\mathrm{median}(|\hat{u}_{\theta,k}-\hat{u}_k|^2)$. All
    spectral models, except the richer spectral model with
    $\epsilon>0.5$, accurately recover dominant modes and suppress
    spurious high-frequency components.}
  \label{fig:absolute_error}
\end{figure}

Figure~\ref{fig:absolute_error} shows that spectral models reconstruct low-frequency modes accurately while suppressing high-frequency noise.
HEA circuits, in contrast, allocate capacity to less relevant modes.

\subsection{Gradient Behavior}

\begin{figure}[t]
  \centering
  \includegraphics[width=\linewidth]{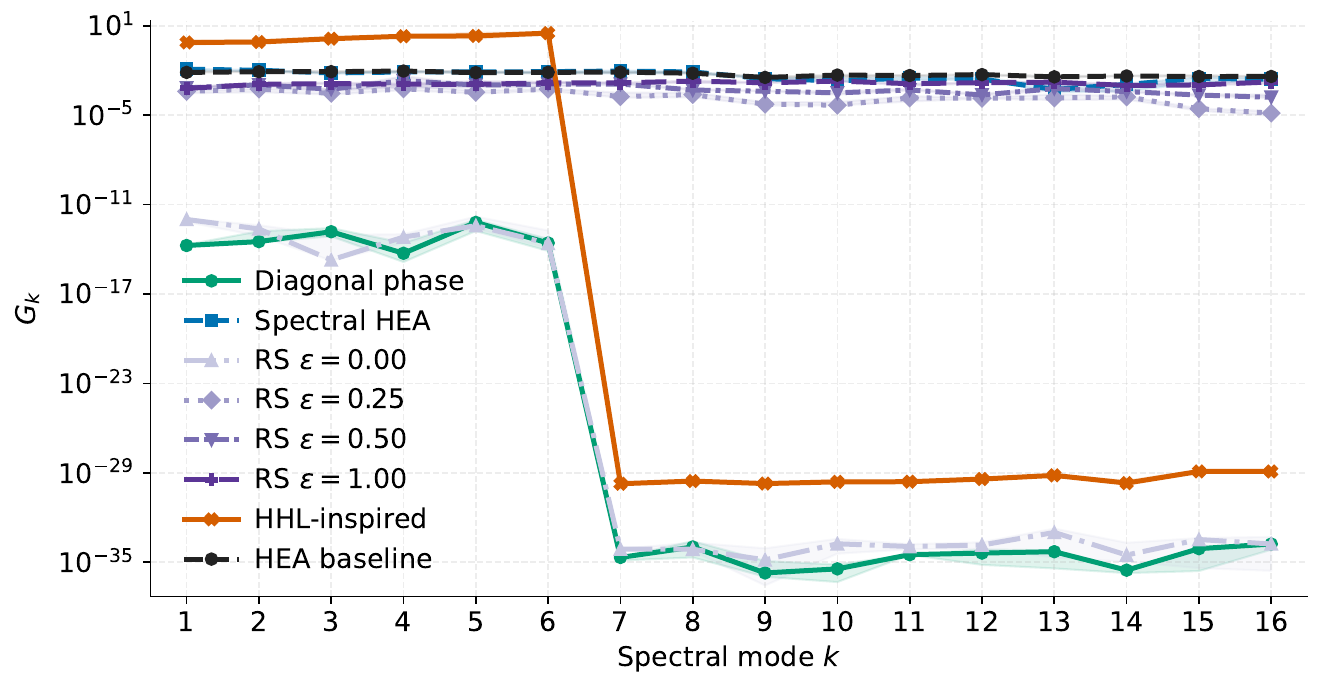}
  \caption{Mode-wise gradient power $G_k$. All spectral models,
    except the richer spectral model with $\epsilon>0.5$, exhibit
    better sensitivity on the gradients of modes of the true solution.}
  \label{fig:gradient_power}
\end{figure}

\begin{figure}[t]
  \centering
  \includegraphics[width=\linewidth]{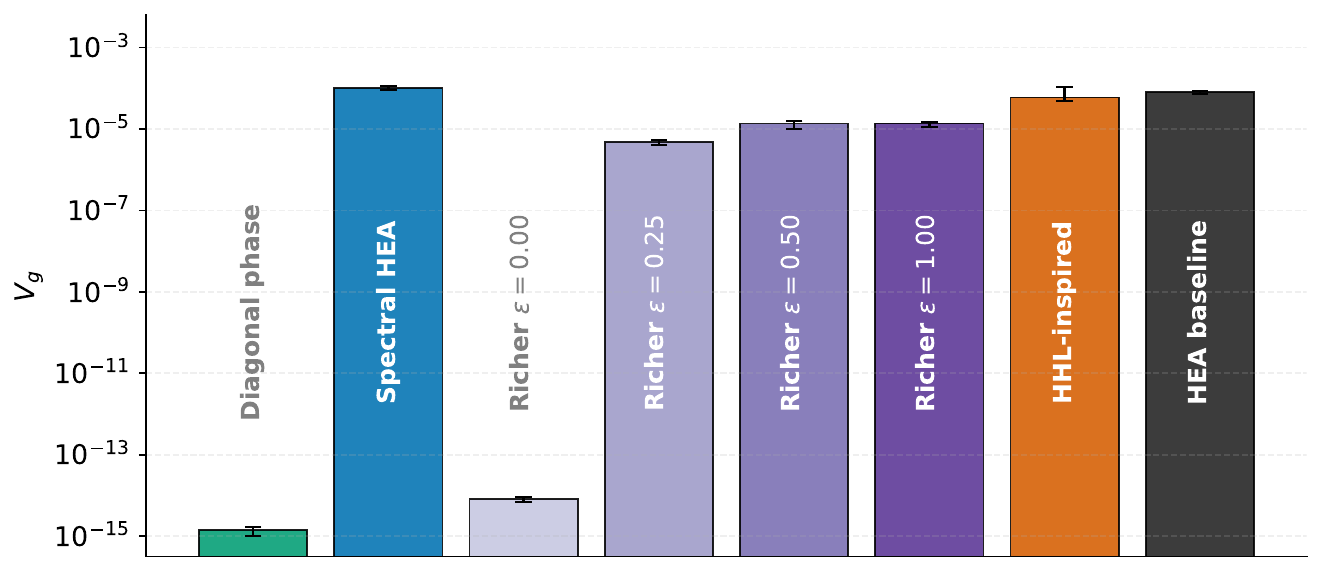}
  \caption{Gradient variance across parameters. Extremely small
    values could indicate flat optimization landscapes.}
  \label{fig:grad_variance}
\end{figure}

Figure~\ref{fig:gradient_power} shows that QSMs concentrate gradient signal on relevant modes, improving parameter efficiency.
Figure~\ref{fig:grad_variance} reveals that gradient magnitude alone does not determine trainability: the diagonal-phase model converges well despite low variance, while the HHL-inspired model matches baseline variance but achieves better accuracy and convergence.
This suggests that gradient structure and alignment are more important than magnitude alone.

\subsection{Expressibility}

\begin{figure}[t]
  \centering
  \includegraphics[width=\linewidth]{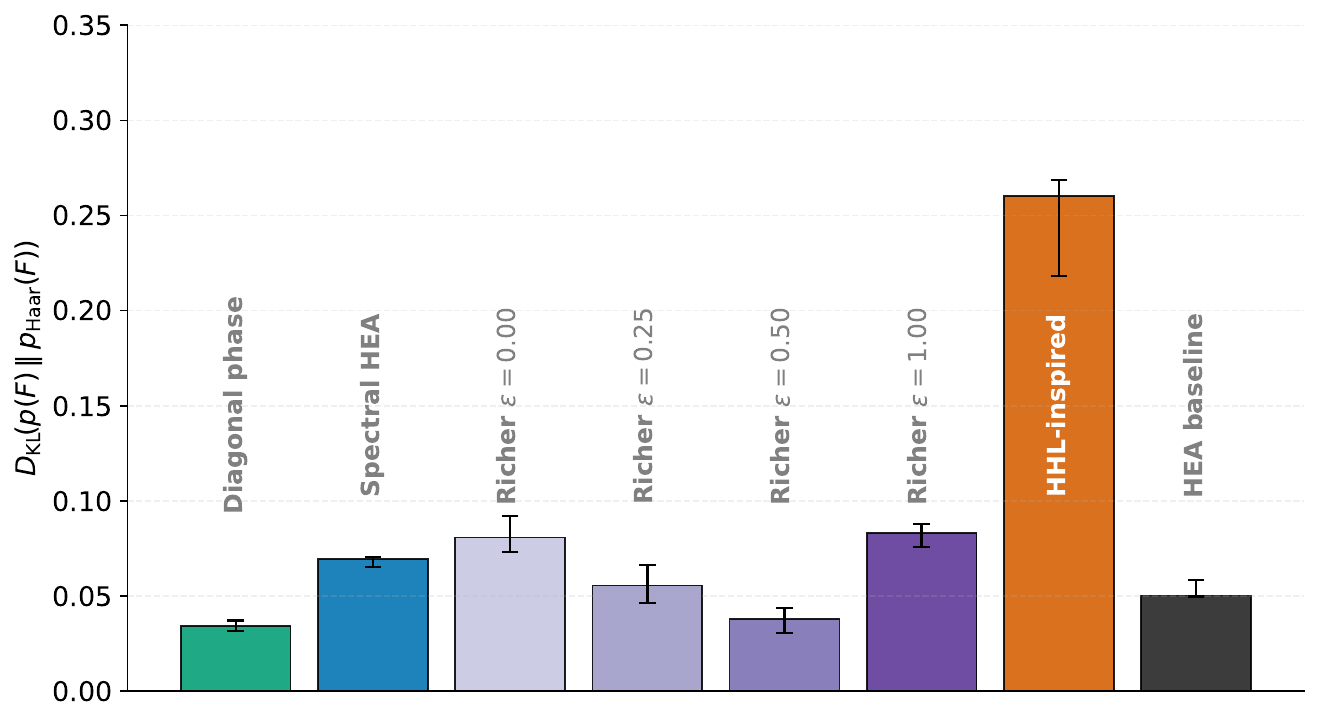}
  \caption{Expressibility measured via KL divergence to Haar fidelity
    distribution. Lower values indicate higher expressibility.}
  \label{fig:expressibility}
\end{figure}

Figure~\ref{fig:expressibility} shows that higher expressibility does not imply better performance.

Moderately expressive spectral models often achieve faster convergence and lower error.
The HHL-inspired model, despite low expressibility, maintains strong gradient behavior and accuracy, indicating that operator-aligned structure is more important than unrestricted expressibility.

\subsection{Approximate Bases and Architecture Comparison}

Even when the basis does not diagonalize the operator, spectral models remain competitive and often outperform baselines, showing robustness to imperfect prior knowledge.
Among architectures, HHL-inspired models provide the best balance between convergence, gradient quality, and spectral accuracy, while richer spectral models reveal a complexity threshold beyond which additional expressibility yields diminishing returns.
\subsection{Summary of Empirical Findings}

Across all benchmarks, the experiments consistently show that structured spectral models outperform generic HEA circuits, with faster convergence driven by operator-aligned parameterizations.
Importantly, these advantages do not require exact diagonalization, as approximate bases already provide significant benefits.
Among the architectures considered, HHL-inspired models offer a particularly favorable balance between simplicity and performance.

Together, these results support the central hypothesis of this work: operator-aware structure can be more valuable than unrestricted expressibility when training quantum models for scientific computing.

\section{Discussion}

The results suggest that the choice of basis representation can be as important as the circuit architecture when designing trainable quantum models.
For linear PDEs, Fourier- or sine-type transforms concentrate the relevant solution structure into a compact set of modes.
Restricting trainable operations to this transformed representation reduces the effective search space and biases the model toward physically meaningful spectra, explaining the improved convergence and spectral reconstruction observed for QSMs.

A key finding is that exact diagonalization is not required.
In the variable-coefficient Poisson problem, the sine basis no longer diagonalizes the operator, yet structured spectral models remain competitive and often outperform HEA baselines.
This suggests that approximate or physically motivated bases can still expose useful low-complexity structure, which is important for realistic systems with heterogeneous coefficients or incomplete physical knowledge.

The richer spectral family also suggests an intermediate complexity regime.
Models with moderate mixing strength, typically $\epsilon\leq 0.5$, often outperform more strongly mixed variants in loss convergence and spectral reconstruction, despite having similar expressibility and average gradient variance.

This indicates that increasing circuit complexity beyond a threshold can yield diminishing or adverse returns.
Analytically characterizing this transition is an important direction for future work.

Finally, some QSM architectures may admit efficient classical approximations because they restrict the trainable operator to structured subspaces.
Thus, potential quantum advantage should be sought in regimes where structured transforms, state preparation, and measurement remain favorable on quantum hardware.
Extending QSMs to learned or data-adaptive bases, for example in image, graph, or wavelet representations, is another promising direction.

\section{Conclusion}

We introduced QSMs, trainable quantum architectures that learn solution operators in transformed coordinates rather than directly in the computational basis.
Across Poisson, Helmholtz, and variable-coefficient Poisson benchmarks, QSMs produced more accurate spectral reconstructions and faster convergence than generic HEA circuits.
They suppressed spurious high-frequency components while preserving physically relevant modes.

Our results show that higher expressibility does not necessarily imply better performance.
Structured models, especially HHL-inspired and moderately mixed spectral architectures, often achieved better convergence, gradient behavior, and spectral accuracy than more expressive alternatives.
Importantly, these advantages persisted even when the chosen basis did not exactly diagonalize the operator.
This supports the view that operator-aware spectral structure can be more valuable than unrestricted expressibility for trainable quantum scientific computing.

\textbf{Code availability.} The code and data used in this work are
publicly available at
\href{https://github.com/cirKITers/Trainable-Quantum-Spectral-Models-for-Partial-Differential-Equations}{Github}.

\clearpage

\bibliographystyle{IEEEtran}
\bibliography{references}

\end{document}